\def\ap3m{AP$^3$M}
\def\LCDM{$\Lambda$CDM}
\def\LWDM{$\Lambda$WDM}

\def\hMpc{$h^{-1}{\ }{\rm Mpc}$}

\def\hMsun{$h^{-1}{\ }{\rm M_{\odot}}$}

\def\nbody{$N$-body}
\def\c15{$c_{\rm 1/5}$}

\newcommand{\Fig}[1]{Fig.~\ref{#1}}
\newcommand{\Table}[1]{Table~\ref{#1}}

\def\ea{et~al.~}                            

\newcommand{\ApJ}[3]    {\mbox{ApJ~\textbf{#1},~#2~(#3)}}

\newcommand{\ApJL}[3]   {\mbox{ApJ~Lett.~\textbf{#1},~#2~(#3)}}

\newcommand{\MNRAS}[3]  {\mbox{MNRAS~\textbf{#1},~#2~(#3)}}

\documentclass{kluwer}    
\usepackage[dvips]{graphicx}

\newdisplay{guess}{Conjecture}

\begin{document}                                                                                   
\begin{article}
\begin{opening}         
\title{Non-Standard Structure Formation Scenarios} 
\author{Alexander \surname{Knebe}$^1$, Brett \surname{Little}$^{1,2}$, Ranty \surname{Islam}$^3$, Julien \surname{Devriendt}$^3$, Asim \surname{Mahmood}$^3$, Joe \surname{Silk}$^3$}  
\runningauthor{Knebe~\ea}
\runningtitle{Non-Standard Structure Formation Scenarios}
\institute{$^1$Centre for Astrophysics~\& Supercomputing, Swinburne University, 
               Australia\\
           $^2$Australian National University, Australia\\
           $^3$Denys Wilkinson Building, Keble Road, Oxford, OX1 3RH, UK}

\begin{abstract}
Observations on galactic scales seem to be in contradiction with
recent high resolution \nbody\ simulations. This so-called cold dark
matter (CDM) crisis has been addressed in several ways, ranging from a
change in fundamental physics by introducing self-interacting cold
dark matter particles to a tuning of complex astrophysical processes
such as global and/or local feedback.  All these efforts attempt to
soften density profiles and reduce the abundance of satellites in
simulated galaxy halos.  In this contribution we are exploring the
differences between a Warm Dark Matter model and a CDM model where the
power on a certain scale is reduced by introducing a narrow negative
feature (''dip''). This dip is placed in a way so as to mimic the loss
of power in the WDM model: both models have the same integrated power
out to the scale where the power of the Dip model rises to the level
of the unperturbed CDM spectrum again. Using \nbody\ simulations we
show that that the new Dip model appears to be a viable alternative to
WDM while being based on different physics: where WDM requires the
introduction of a new particle species the Dip stems from a
non-standard inflationary period. If we are looking for an alternative
to the currently challenged standard \LCDM\ structure formation
scenario, neither the \LWDM\ nor the new Dip model can be ruled out
with respect to the analysis presented in this contribution. They both
make very similar predictions and the degeneracy between them can only
be broken with observations yet to come.
\end{abstract}

\end{opening}           

\section{The Setup}  

The so-called Cold Dark Matter crisis has led to a vast number of
publications trying to solve the problems which seem to be associated
with an excess of power on small scales. One possibility to reduce
this power is to introduce Warm Dark Matter (i.e. Knebe~\ea 2002;
Bode, Ostriker~\& Turok 2001; Avila-Reese~\ea 2001; Colin~\ea 2000).
But another way to decrease power on a certain scale is to introduce a
negative feature (``dip'') into an otherwise unperturbed CDM power
spectrum (cf. Knebe~\ea 2001). Several mechanisms have been proposed
that could generate such features in the primordial spectrum during
the epoch of inflation. Among these are models with broken scale
invariance (BSI) (Lesgourgues, Polarski~\& Starobinsky 1998), and
particularly BSI due to phase transitions during inflation
(Barriga~\ea 2000).

The \LWDM\ and the fiducial \LCDM\ model used in this paper are the
same as those presented in Knebe~\ea (2002) with the cosmological
parameters $\Omega_0 = 1/3$, $\lambda_0 = 2/3$, $\sigma_8=0.88$,
$h=2/3$, and $m_{\rm WDM} = 0.5$keV for \LWDM. For the Dip model we
are using the same prescription to introduce a Gaussian feature into
an otherwise unperturbed CDM power spectrum as outlined in Knebe,
Islam~\& Silk (2001) with the parameters $A$=-0.995, $\sigma_{\rm
mod}$=0.5, and $2\pi/k_0$=1.8\hMpc.

\begin{figure}
\centerline{\includegraphics[width=8.7cm]{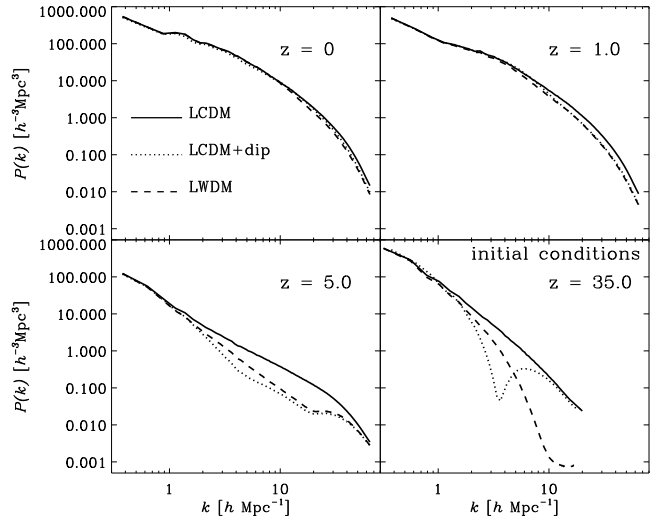}}
\caption{Evolution of the power spectrum as measured on a regular 512$^3$ grid.}
\label{power}
\end{figure}

\section{The Outcome}
In \Fig{power} we show the evolution of the dark matter power
spectrum. We clearly see that the features are well represented in the
initial conditions. But it is important to note that the non-linear
transfer of power from large to small scales has washed out these
features completely by redshift $z$=0. Therefore it is impossible to
use current observations of the (galaxy) power spectrum as for
instance measured by the 2dF team (Percival~\ea 2001) to set
constraints on features on such small scales (e.g. 1.8\hMpc) in the
primordial power spectrum.

\begin{table}
\caption{Number of satellite galaxies within the virial radius 
         of the most massive halo.}
\label{satellites}
\begin{tabular}{lccc}
\hline
 redshift  & \LCDM & \LCDM+Dip & \LWDM \\
\hline
 $z=0$     &   42  &    30     &   29  \\
 $z=1$     &   29  &    29     &   29  \\
\hline
\end{tabular}
\end{table}

\begin{figure}
\centerline{\includegraphics[width=8.7cm]{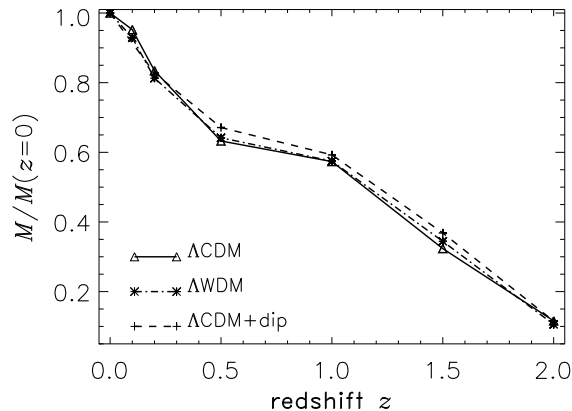}}
\caption{Evolution of the mass of the most massive halo in all three models.}
\label{masshistory}
\end{figure}

One of the major problems with Cold Dark Matter is the over-prediction
of satellite galaxies orbiting within a galactic halo (Klypin \ea
1999, Moore \ea 1999). In \Table{satellites} we summarize the number
of satellites found in the most massive dark matter halo. It indicates
that both non-standard models are able to overcome that
problem. However, \Fig{masshistory} shows that the mass history of
that halo is indistinguishable for all three models and hence the
differences for $z=0$ (along with the agreement for $z=1$) in
\Table{satellites} can now only be explained by a different ratio of
satellite accretion to satellite destruction throughout the models; in
our non-standard models satellite galaxies are more easily disrupted
because they are less concentrated.

\begin{figure}
\centerline{\includegraphics[width=8.7cm]{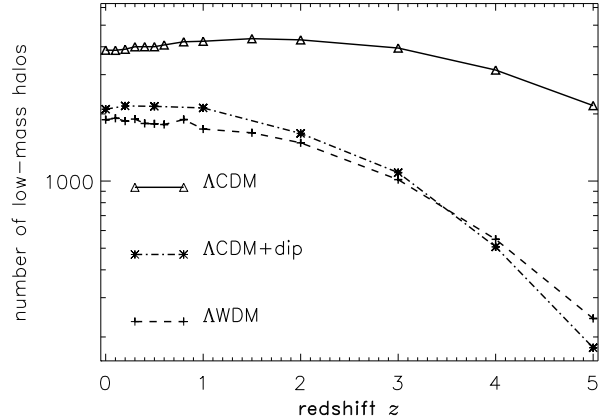}}
\caption{Abundance evolution of objects in the mass range 
         $10^{10}$\hMsun $< M < 10^{11}$\hMsun.}
\label{abundance}
\end{figure}

Another interesting outcome is the formation history as well as the
formation sites of low-mass galaxies $M\in
[10^{10},10^{11}]$\hMsun. \Fig{abundance} demonstrates that such halos
do form at later times compared to the standard \LCDM\ model and a
detailed analysis of their formation sites shows that they are
preferably forming along the filaments.

\section{The Conclusions}
We conclude from the given analysis that both our non-standard models
work equally well even though they are based on completely different
physical processes; where WDM requires the introduction of a new
particle species the Dip model is based on a non-standard inflationary
period.  The only way of breaking their degeneracy might lie within
the filamentary structures of the Universe and the low-mass end of the
mass function. But to strengthen these findings we need more detailed
and much better resolved simulations in the future.


\acknowledgements The simulations were run at the Centre for 
Astrophysics~\& Supercomputing, Swinburne University and the Oxford
Supercomputer Centre. We are grateful to A. Klypin and A. Kravtsov for
kindly providing a copy of the ART code that was used for the
\LCDM\ and \LWDM\ simulation, respectively.


\end{article}

\begin{thebibliography}{}

\bibitem[avila01]{avila01}
        {Avila-Reese V., Colin P., Valenzuela O., D'Onghia E., Firmani C.,
         \ApJ{559}{516}{2001}}

\bibitem[barriga00]{barriga00}
        {Barriga J., Gazta\~naga E., Santos M.G, Sarkar S. \MNRAS{324}{977}{2001}}

\bibitem[bode01]{bode01}
        {Bode P., Ostriker J.P., Turok N., \ApJ{556}{93}{2001}}

\bibitem[colin00]{colin00}
        {Colin P., Avila-Reese V., Valenzuela O., \ApJ{542}{622}{2000}}

\bibitem[klypin99]{klypin99}
        {Klypin A.A., Kravtsov A.V., Valenzuela O., Prada F.,
         \ApJ{522}{82}{1999}}

\bibitem[knebe01]{knebe01}
        {Knebe A., Islam R.R., Silk J., \MNRAS{326}{109}{2001}}

\bibitem[knebe02]{knebe02}
        {Knebe A., Devriendt J., Mahmood A., Silk J.,\MNRAS{329}{813}{2002}}

\bibitem[lesgourges]{lesgourges}
	{Lesgourgues J., Polarski D., Starobinsky A.A.,\MNRAS{297}{769L}{1998}}

\bibitem[moore99]{moore99}
        {Moore B., Ghigna S., Governato F., Lake G., Quinn T., Stadel J.,
         Tozzi P., \ApJL{524}{19}{1999}}

\bibitem[percival]{percival}
	{Percival W.J.~\ea,\MNRAS{327}{1297}{2001}}


\end{thebibliography}
\end{document}